# Carrier conversion from terahertz wave to dual-wavelength near-infrared light injection-locking to optical comb using asynchronous nonpolarimetric electro-optic downconversion with electro-optic polymer modulator

Yudai Matsumura[1], Yu Tokizane[2*], Eiji Hase[2], Naoya Kuse[2,3], Takeo Minamikawa[2], Jun-ichi Fujikata[2], Hiroki Kishikawa[2], Masanobu Haraguchi[2], Yasuhiro Okamura[2,4], Takahiro Kaji[5], Akira Otomo[5], Isao Morohashi[5], Atsushi Kanno[5,6], Shintaro Hisatake[7], and Takeshi Yasui[2*]

[1]Graduate School of Sciences and Technology for Innovation, Tokushima University, Tokushima, Tokushima 770-8506, Japan

[2]Institute of Post-LED Photonics (pLED), Tokushima University, Tokushima, Tokushima 770-8506, Japan

[3]PRESTO, Japan Science and Technology Agency, Kawaguchi, Saitama, 332-0012, Japan

[4]Center for Higher Education and Digital Transformation, University of Yamanashi, Kofu, Yamanashi 400-8510, Japan

[5]National Institute of Information and Communications Technology, Koganei, Tokyo 184-8795, Japan

[6]Department of Electrical and Mechanical Engineering, Nagoya Institute of Technology, Nagoya, Aichi 466-8555, Japan

[7]Electrical and Energy System Engineering Division, Gifu University, Gifu, Gifu 501-1193, Japan

E-mail: tokizane@tokushima-u.ac.jp, yasui.takeshi@tokushima-u.ac.jp

THz waves are promising wireless carriers for next-generation wireless communications, where a seamless connection from wireless to optical communication is required. In this study, we demonstrate carrier conversion from THz waves to dual-wavelength NIR light injection-locking to an optical frequency comb using asynchronous nonpolarimetric electro-optic downconversion with an electro-optic polymer modulator. THz wave in the W band was obtained as a stable photonic RF beat signal of 1 GHz with a signal-to-noise ratio of 25 dB via the proposed THz-to-NIR carrier conversion. In addition, the results imply the potential of the photonic detection of THz waves for wireless-to-optical seamless communication.

Terahertz (THz) waves are wireless carriers for next-generation mobile communications (6G, expected carrier frequency > 300 GHz).[1] Despite wireless electronics being widely used until recent wireless communications (5G, carrier frequency = 28 GHz or more), they may face technical limitations in 6G owing to the high frequency. For example, increased phase noise of wireless carriers and/or increased signal loss in electric transmission lines. One potential method to overcome this technical limitation is to use photonic technology in THz communications.[2] A promising approach for the photonic generation of low-phase-noise THz waves (freq. = $f_{THz}$) is the use of an optical frequency comb (OFC)[3-6] owing to the low phase noise of frequency spacing (= $f_{rep}$), where two OFC modes with THz frequency spacing (= $mf_{rep} = f_{THz}$) are extracted and used for photomixing with the help of a uni-traveling carrier photodiode (UTC-PD).[7,8] Particularly, on-chip Kerr microresonator soliton combs with a considerably large $f_{rep}$, namely, the soliton microcomb, have attracted attention for the photonic generation of ultralow-phase-noise THz waves at 300 GHz[9] and 560 GHz[10] because $f_{rep}$ is close to $f_{THz}$. Thus, two adjacent microcomb modes can be directly used for photomixing, benefiting from the low-phase-noise advantage of $f_{rep}$ without the influence of optical frequency multiplication. Furthermore, photonic-generated THz waves have been applied to wireless data transfer.[11,12] However, these demonstrations still use electronic THz detectors, such as Schottky barrier diodes and fundamental mixers, which are bulky, fragile, complicated, and expensive.

If carrier conversion from THz waves to near-infrared (NIR) light can be realized by use of photonic technology, existing mature photonic devices for optical communication can be used without minimal modification, enabling photonic THz detection for a seamless connection from wireless communication to optical communication. One potential method for such THz-to-NIR carrier conversion is the use of nonpolarimetric electro-optic (EO) downconversion (NP-EO-DC), which has been applied for the visualization of millimeter-wave electric fields.[13-16] In this method, dual-wavelength NIR light is used as an optical carrier, and millimeter-wave-induced modulation sidebands are generated corresponding to to the dual-wavelength NIR light via the EO effect. When the optical frequency spacing of the dual-wavelength NIR light is close to the millimeter-wave frequency, an optical beat signal between one of the dual-wavelength NIR light beams and the modulation sideband of the other appears in the RF region. The heterodyned detection of the optical beat signal enhances the RF beat signal corresponding to the baseband signal. In particular, since the asynchronous NP-EO-DC[15,16] uses two OFC modes with a frequency spacing close to the frequency of electromagnetic waves being extracted from an electro-optic modulator OFC (EOM-OFC) for stable dual-wavelength NIR light, it is applicable for self-oscillating electromagnetic wave and thus has a good affinity with THz-to-NIR carrier conversion. The remaining bottleneck is the limited OSNR of this dual-wavelength NIR light owing to the background of residual unwanted EOM-OFC modes and/or amplified spontaneous emission (ASE). The means to eliminate this bottleneck is injection-locking of high-power CW lasers to EOM-OFC modes, enabling the amplification and phase noise transfer of each EOM-OFC mode while eliminating the residual unwanted EOM-OFC modes and reducing the ASE background. Although such dual-wavelength NIR light injection-locking to EOM-OFC modes have effectively applied for photonic generation of millimeter-wave and THz waves with UTC-PD,[17,18] there have been no attempts to apply it for the asynchronous NP-EO-DC.

In this study, for a proof-of-concept experiment prior to the demonstration at the 6G carrier frequency band, we demonstrate THz-to-NIR carrier conversion in the W-band by the asynchronous NP-EO-DC using high-OSNR, stable, dual-wavelength NIR light injection-locking to a 10-GHz-spacing EOM-OFC[19,20]). We simultaneously use an electro-optic polymer (EOP) modulator,[21,22] in place of EO crystals used in previous research, because it has advantages such as high-frequency response, long interaction length, low absorption of THz waves, a high EO coefficient, and good phase matching between THz and optical carriers.

Figure 1(a) shows the principle of carrier conversion from a modulated THz wave (wireless carrier, carrier freq. = $f_{THz}$, modulation width = $\Delta f_{THz}$) to dual-wavelength NIR light (optical carrier, opt. freq. = $\nu_1$, $\nu_2$). The frequency spacing between the pair of high-power CW lasers (CWL1, opt. freq. = $\nu_1$; CWL2, opt. freq. = $\nu_2$; opt. freq. spacing = $\nu_2 - \nu_1$) was set close to $f_{THz}$. When two OFC modes (opt. freq. = $\nu_1$, $\nu_2$) with a frequency spacing of $mf_{rep}$ ($\approx \nu_2 - \nu_1 = f_{THz}$), extracted by an optical bandpass filter (BPF), is respectively incident into two CW lasers, the injection locking of each CW laser to each OFC mode enables amplification and phase noise transfer of two OFC modes while eliminating the residual unwanted EOM-OFC modes and reducing the ASE background. The resulting stable dual-wavelength NIR light (i.e., $\nu_1$ carrier and $\nu_2$ carrier) was used for the THz-to-NIR carrier conversion. Thereafter, the modulated THz wave is incident onto an antenna of an EOP modulator as an external electric field, whereas $\nu_1$ carrier propagates through an optical waveguide in the EOP modulator as optical carriers. The phase modulation in the EOP modulator creates the modulation sidebands of $\nu_1$ carrier that are only $f_{THz}$ apart. $\nu_2$ carrier bypasses the EOP modulator for an unmodulated optical carrier. As the frequency difference between $\nu_1$ carrier and $\nu_2$ one (= $\nu_2 - \nu_1 = mf_{rep}$) is set to be close to $f_{THz}$, $\nu_1$ sideband appears near $\nu_2$ carrier. Importantly, the $\nu_1$ and $\nu_2$ carriers are phase-locked to each other because of their injection-locking to the EOM-OFC modes, and their fluctuations in optical frequency are common to each other while maintaining a constant frequency spacing of $mf_{rep}$. Therefore, a pair of $\nu_2$ carriers and $\nu_1$ sidebands is equivalent to a pair of $\nu_1$ carriers and $\nu_1$ sidebands except their frequency difference. An optical beat signal between the $\nu_2$ carrier and $\nu_1$ sideband was detected as a baseband signal by the photodetector after optical bandpass filtering. The use of optical heterodyning detection enables us to enhance the weak modulation of the sideband signal. The resulting baseband signal (freq. $= f_{RF} = f_{THz} - mf_{rep}$) corresponded to the modulation signal of the THz carrier wave.

Figure 1(b) shows a schematic of the experimental setup. We used a pair of 4-mW distributed feedback laser diodes (DFB1,Gooch & Housego, AA1408-193350-100-PM900-FCA-NA, $\nu_1 = 193.39$ THz corresponding to a wavelength of 1550.2 nm; DFB2, Gooch & Housego, AA1408-193350-100-PM900-FCA-NA, $\nu_2 = 193.49$THz corresponding to a wavelength of 1549.4 nm) for slave lasers and a pair of 2-µW EOM-OFC modes for master lasers, respectively. Two EOM-OFC modes with a frequency spacing of $10f_{rep}$ (EOM-OFC-M1 and EOM-OFC-M2) were extracted from an EOM-OFC (center wavelength = 1550 nm, $f_{rep} = 10$ GHz)[19,20] by tunable ultra-narrowband optical bandpass filters (BPFs; Alnair labs, CVF-300CL-PM-FA, center wavelength = 1525~1610 nm, optical passband = 3.7~370 GHz corresponding to 20~3000 pm, insertion loss = 5.5 dB). The EOM-OFC-M1 and EOM-OFC-M2 were incident on DFB1 and DFB2, respectively, via the circulators. When $\nu_1$ and $\nu_2$ were detuned to the optical frequencies of EOM-OFC-M1 and EOM-OFC-M2 within the locking range of DFB1 and DFB2 (typically a few hundred MHz) by current control, injection-locking was achieved. The details of injection locking are given elsewhere.[17,18]

The $\nu_1$ carrier injection locking to EOM-OFC-M1 was fed into the optical waveguide of a non-coplanar patch-antenna-type EOP modulator[22] whereas the $\nu_2$ carrier injection-locking to EOM-OFC-M2 bypassed the EOP modulator. When THz wave ($f_{THz} = 101$ GHz, power = 4 dBm = 2.5 mW) was irradiated as an elliptic focus on the parch antenna by a combination of a spherical THz lens (L1, diameter = 50 mm, focal length = 100 mm) and a crossed pair of cylindrical THz lenses (CL1, size = 30 mm by 50 mm, focal length = 96 mm; CL2, size = 85 mm by 85 mm, focal length = 50 mm). This results in the generation of an $\nu_1$ sideband separated by $f_{THz}$ from the $\nu_1$ carrier. The optical beat signal between the $\nu_1$ sideband and the $\nu_2$ carrier was generated by interfering with them via a fiber coupler and then detected by a photodiode (PD, wavelength = 1550 nm, RF bandwidth = 10 GHz) after passing through another BPF. The RF beat signal was acquired by an electric spectrum analyzer (ESA, RIGOL, DSA815, freq. = 9kHz-1.5GHz).

First, we evaluated the basic performance of injection locking of the DFB to the EOM-OFC modes. The red plot in Fig. 2(a) shows the optical spectrum of the $\nu_I$ carrier (optical power = 8 mW) injection-locking to EOM-OFC-M1 (resolution = 0.02 nm). For comparison, the blue plot shows the optical spectrum of EOM-OFC-M1 (optical power = 2 μW) extracted from EOM-OFC. The wavelength-magnified optical spectrum of them is shown in Fig. 2(b). The wavelength of the injection-locked $\nu_I$ carrier was the same as that of the extracted EOM-OFC-M1, indicating successful injection locking. Despite ultrafine spectrally modulated luminescence owing to the internal grating structure characteristic of the DFB, which exists in both tails of the injection-locking $\nu_I$ carrier, its intensity is sufficiently lower than that of the the $\nu_I$ carrier, and the resulting OSNR is to 70 dB. However, the OSNR of the extracted EOM-OFC-M1 remained at approximately 50 dB because of the limited optical power of each EOM-OFC mode and the noise background. Thus, we obtained a sufficient OSNR in the $\nu_I$ carrier for the generation of modulation sidebands using the EOP modulator.

Subsequently, the modulation sideband of the $\nu_I$ carrier was observed by irradiating the EOP modulator with THz waves. Figures 3(a) and 3(b) compare the optical spectra of the output light from the optical waveguide of the EOP modulator with and without THz-wave irradiation (resolution = 0.02 nm). Both the $\nu_I$ carrier and its modulation sideband are observed under THz irradiation, whereas only the $\nu_I$ carrier is observed without irradiation. The $\nu_I$ modulation sideband appeared exactly $\pm f_{THz}$ (= 101 GHz) away from the $\nu_I$ carrier. The CSR is 40 dB for $\nu_I$ carrier and $\nu_I$ modulation sideband. The improved CSR results from shaping the THz focus as an oval to match the shape of the patch antenna. The OSNR of the $\nu_I$ modulation sideband was clearly confirmed at 30 dB, owing to the enhanced OSNR of the $\nu_I$ carrier by injection locking. Similar CSR and OSNR should be maintained for a pair of $\nu_2$ carrier and $\nu_I$ modulation sideband that are close to each other.

Finally, we measured the RF spectrum of the optical beat signal between the $\nu_2$ carrier and $\nu_I$ modulation sidebands. Figure 4 shows the RF spectrum of the beat signal (resolution = 1 MHz). Because the frequency difference between $f_{THz}$ (= 101 GHz) and $10f_{rep}$ (= 100 GHz) was 1 GHz, the RF beat signal between them was observed at 1 GHz, achieving an SNR of 25 dB. To confirm the effectiveness of the stable dual-wavelength NIR light injection-locking to the two EOM-OFC modes in the present THz-to-NIR carrier conversion, we continuously monitored the behavior of the RF spectrum before and after the injection locking was lost, as shown in Movie 1. With active injection locking, the RF beat signal remained fixed at 1 GHz. Because the RF beat signal is sufficiently stable in frequency and phase owing to injection locking in the EOM-OFC, it can be used for THz wireless communication. However, once the injection locking was inactive, the RF beat signal fluctuated significantly and/or multiple RF beat signals appeared, making it difficult to apply the RF beat signal to THz wireless communication.

We used dual-wavelength NIR light injection-locking to the EOM-OFC modes as an optical carrier in the asynchronous NP-EO-DC using the EOP modulator. We here discuss the effectiveness of the injection locking in the asynchronous NP-EO-DC by comparing optical spectra of $\nu$ carrier and its modulation sidebands when using the DFB1 injection-locking to EOM-OFC-M1 and the extracted EOM-OFC-M1 as a dual-wavelength NIR light, respectively. We set the optical power of them fed into the EOP modulator to be 3.3 mW, which was limited by the maximum optical power of the extracted EOM-OFC-M1. Figure 5(a) compares optical spectra of the DFB1 injection-locking to EOM-OFC-M1 without and with irradiation of THz wave (resolution bandwidth = 0.02 nm). The $\nu$ modulation sidebands clearly appeared without interference of residual unwanted EOM-OFC modes due to the injection locking. Conversely, Fig. 5(b) shows the comparison of optical spectra of the extracted EOM-OFC-M1 without and with irradiation of THz wave. Even though the $\nu$ modulation sidebands has no modulation bandwidth due to no modulation of THz wave, residual unwanted EOM-OFC modes significantly interfered their modulation sidebands, limiting the OSNR of the modulation sideband signal. If the modulation bandwidth of THz wave becomes wide by coding the modulation signal into THz carrier wave, such interference effect will become large that cannot be ignored. We also compare RF spectra of the optical beat signal generated by the $\nu_2$ carrier and $\nu_1$ modulation sidebands between the DFB1 injection-locking to EOM-OFC-M1 and the extracted EOM-OFC-M1. Figures 5(c) and 5(d) show the RF spectra of the optical beat signal (resolution = 1 MHz) when using the DFB 1 injection-locking to EOM-OFC-M1 (optical power = 3.3 mW) and the extracted EOM-OFC-M1 (optical power = 3.3 mW) as a dual-wavelength NIR light, respectively. Although the RF beat signal was confirmed at 1 GHz in both RF spectra, the SNR was enhanced in the former. In this way, we confirmed the effectiveness of the injection locking in the asynchronous NP-EO-DC.

The remaining technical challenge is further decreasing the relative phase noise in the injection-locking dual-wavelength NIR light and further increasing the measured $f_{THz}$. Despite such injection-locking dual-wavelength NIR light being more stable in frequency and phase than free-running dual-wavelength NIR light, there is still room for a further decrease in its relative phase noise. Particularly, the use of two extracted EOM-OFC modes with a large frequency separation (= $mf_{rep}$) for injection-locking spoils the low phase noise of $f_{rep}$ characteristic in the OFC because it is equivalent to the high-order optical frequency multiplication of $f_{rep}$ and hence increases the relative phase noise of the resulting dual-wavelength NIR light. One promising method for no optical frequency multiplication of $f_{rep}$ is the use of the soliton microcomb instead of the EOM-OFC because it significantly increases $f_{rep}$ up to a few tens of GHz to a few THz, which is much larger than $f_{rep}$ in the EOM-OFC, while achieving stable soliton mode-locking oscillation with low phase noise. This soliton microcomb is a good reference for injection locking without the need for optical frequency multiplication.[23,24] In addition, the use of the soliton microcomb in the asynchronous NP-EO-DC provides a margin to further increase the measured $f_{THz}$ toward the 6G carrier frequency band. Soliton-microcomb-based, asynchronous NP-EO-DC has a good affinity for soliton-microcomb-based photonic THz generation and wireless data transfer at 6G carrier frequency bands of 300 GHz and 560 GHz.[9-12]

In conclusion, we demonstrated THz-to-NIR carrier conversion by combining a high-sensitivity EOP modulator, optical-heterodyne-enhanced detection of the asynchronous NP-EO-DC, and high-OSNR, phase-locked dual-wavelength NIR light injection-locking to the EOM-OFC. The resulting RF beat signal, corresponding to the THz wireless carrier, achieved a SNR of 25 dB and high frequency stability. Despite an unmodulated THz wave being used as the wireless carrier, the achieved SNR and stability of the RF beat signal imply the possibility of extending this method to multilevel modulation using amplitude and/or phase. In our future study, we aim to apply it to wireless data transfer with on-off keying amplitude modulation in the W-band. The proposed method is a powerful tool for photonic THz detection in THz communication.

**Acknowledgments**

This study has been partly conducted under the contract "R&D of high-speed THz communication based on radio and optical direct conversion" (JPJ000254) made with the Ministry of Internal Affairs and Communications of Japan.

Opt. Lett. **19**, 121401 (2021).

## Figure Captions

**Fig. 1.** (a) Principle of carrier conversion from a modulated THz wave to dual-wavelength NIR light. EOM-OFC, electro-optic modulator optical frequency comb; CWL1 and CWL2, high-power CW lasers; EOP modulator, electro-optic polymer modulator. (b) Experimental setup. EOM-OFC, electro-optic modulator optical frequency comb; BPFs, optical bandpass filters; OCs, optical circulators; DFB1 and DFB2, distributed feedback laser diode; POC, polarization controller; EOP-MOD, electro-optic polymer modulator; L1, spherical THz lens; CL1 and CL2, a crossed pair of cylindrical THz lenses; PD, photodiode; ESA, electric spectrum analyzer.

**Fig. 2.** (a) Optical spectra of the $\nu_1$ carrier injection locked to EOM-OFC-M1 (red plot) and EOM-OFC-M1 extracted from EOM-OFC (blue plot) and (b) their magnified spectra. Resolution bandwidth of optical spectrum is 0.02 nm.

**Fig. 3.** (a) Optical spectra of the $\nu_1$ carrier without irradiation of THz wave. (b) Optical spectra of the $\nu_1$ carrier and its modulation sideband with irradiation of THz wave. Resolution bandwidth of optical spectrum is 0.02 nm.

**Fig. 4.** RF spectra of optical beat signal between the $\nu_1$ sideband and the $\nu_2$ carrier. Resolution bandwidth of RF spectrum was 1 MHz.

**Fig. 5.** (a) Optical spectra of the DFB1 injection-locking to EOM-OFC-M1 (optical power = 3.3 mW) without and with irradiation of THz wave. (b) Optical spectra of the EOM-OFC-M1 (optical power = 3.3 mW) without and with irradiation of THz wave. Resolution bandwidth of optical spectrum is 0.02 nm. (c) RF beat spectra obtained by (a) DFB1 injection-locking to EOM-OFC-M1 (optical power = 3.3 mW) and (d) EOM-OFC-M1 (optical power = 3.3 mW). Resolution bandwidth of RF spectrum is 1 MHz.

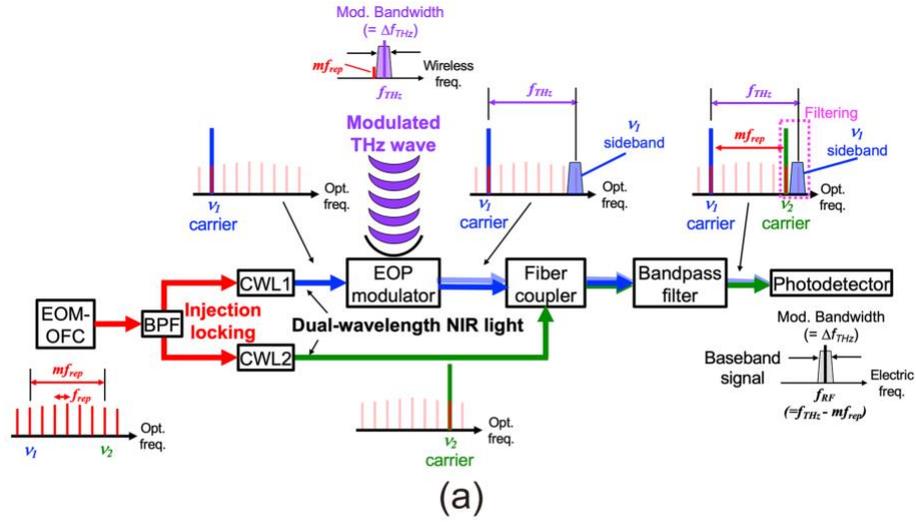

(a)

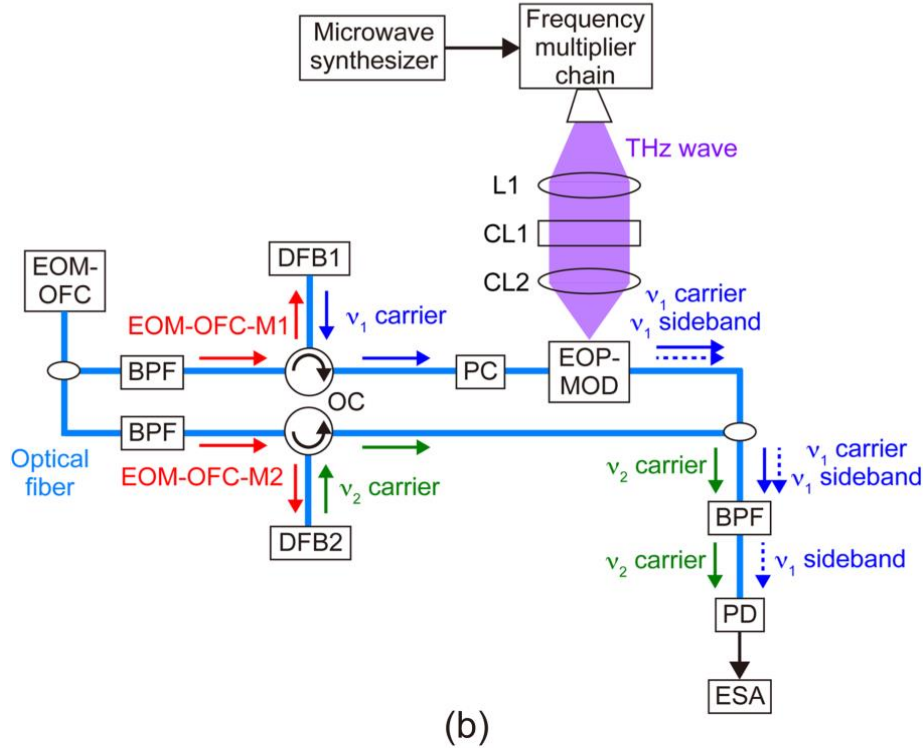

(b)

**Fig. 1.** (a) Principle of carrier conversion from a modulated THz wave to dual-wavelength NIR light. EOM-OFC, electro-optic modulator optical frequency comb; CWL1 and CWL2, high-power CW lasers; EOP modulator, electro-optic polymer modulator. (b) Experimental setup. EOM-OFC, electro-optic modulator optical frequency comb; BPFs, optical bandpass filters; OCs, optical circulators; DFB1 and DFB2, distributed feedback laser diode; POC, polarization controller; EOP-MOD, electro-optic polymer modulator; L1, spherical THz lens; CL1 and CL2, a crossed pair of cylindrical THz lenses; PD, photodiode; ESA, electric spectrum analyzer.

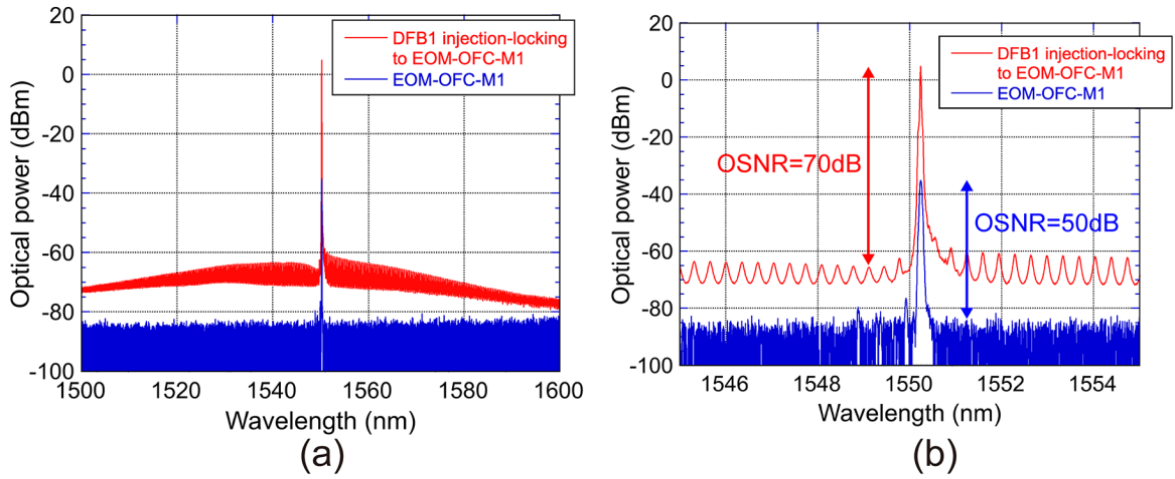

**Fig. 2.** (a) Optical spectra of the $\nu_1$ carrier injection locked to EOM-OFC-M1 (red plot) and EOM-OFC-M1 extracted from EOM-OFC (blue plot) and (b) their magnified spectra. Resolution bandwidth of optical spectrum is 0.02 nm.

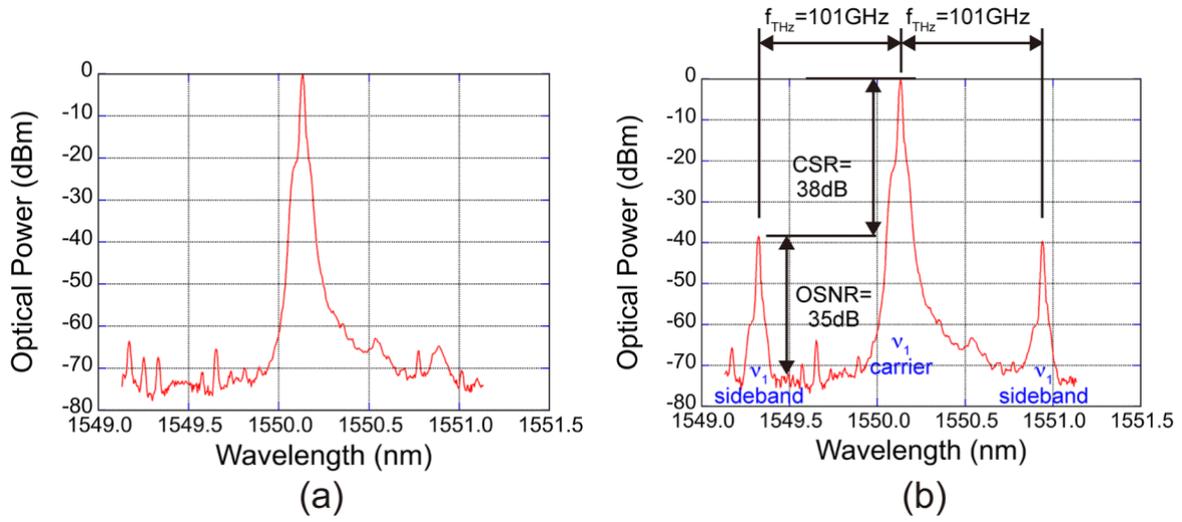

**Fig. 3.** (a) Optical spectra of the $\nu_1$ carrier without irradiation of THz wave. (b) Optical spectra of the $\nu_1$ carrier and its modulation sideband with irradiation of THz wave. Resolution bandwidth of optical spectrum is 0.02 nm.

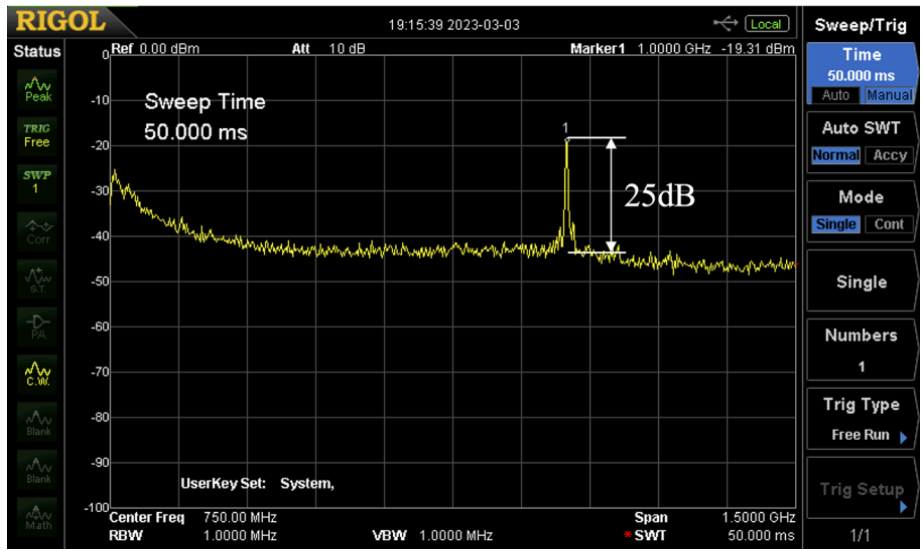

**Fig. 4.** RF spectra of optical beat signal between the $\nu_1$ sideband and the $\nu_2$ carrier. Resolution bandwidth of RF spectrum was 1 MHz.

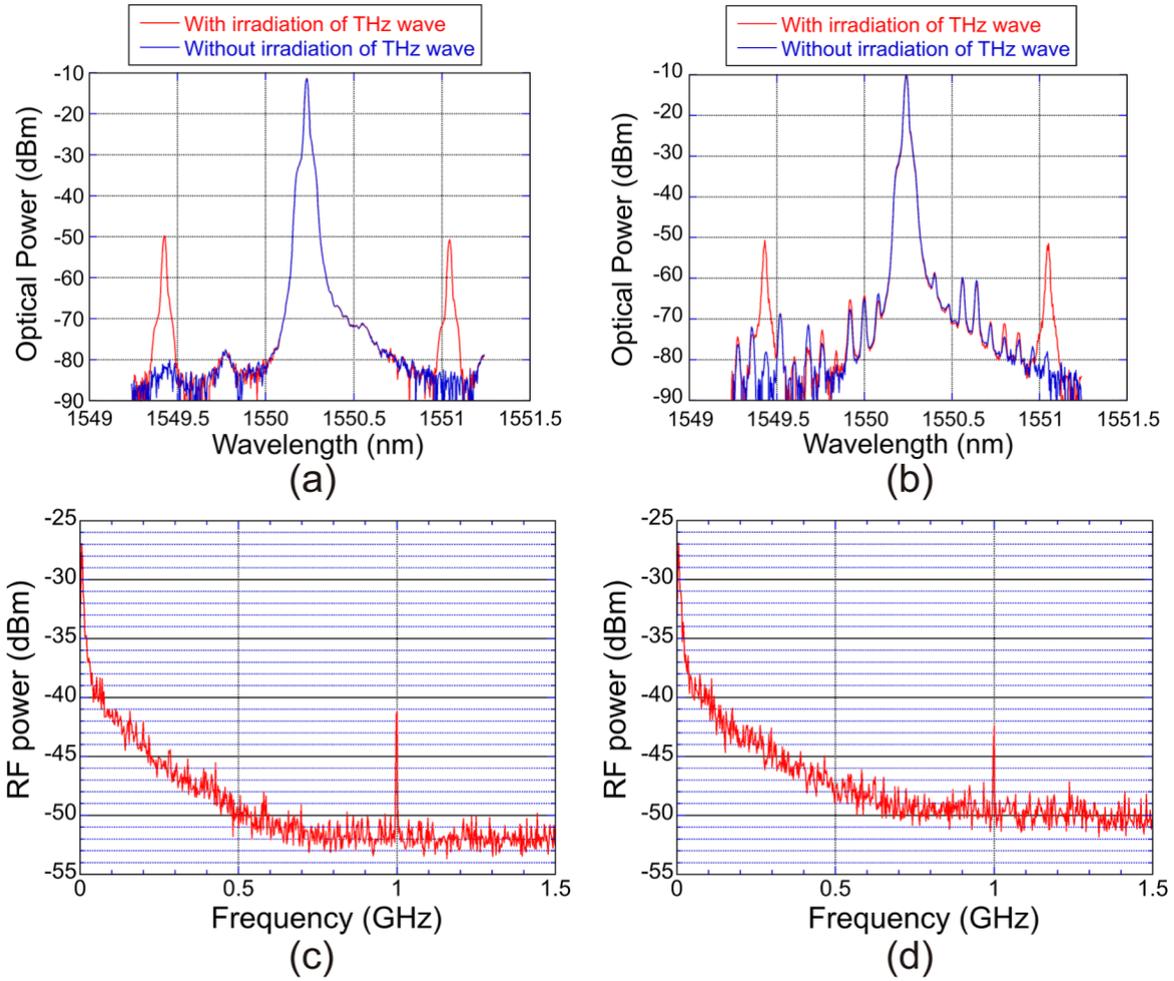

**Fig. 5.** (a) Optical spectra of the DFB1 injection-locking to EOM-OFC-M1 (optical power = 3.3 mW) without and with irradiation of THz wave. (b) Optical spectra of the EOM-OFC-M1 (optical power = 3.3 mW) without and with irradiation of THz wave. Resolution bandwidth of optical spectrum is 0.02 nm. (c) RF beat spectra obtained by (a) DFB1 injection-locking to EOM-OFC-M1 (optical power = 3.3 mW) and (d) EOM-OFC-M1 (optical power = 3.3 mW). Resolution bandwidth of RF spectrum is 1 MHz.